\documentclass[english]{article}
\usepackage[T1]{fontenc}
\usepackage[latin1]{inputenc}
\usepackage{babel}
\makeatletter

\providecommand{\LyX}{L\kern-.1667em\lower.25em\hbox{Y}\kern-.125emX\@}

\usepackage[T1]{fontenc}
\usepackage[latin1]{inputenc}
\usepackage{a4wide}
\date{}

\begin{document}

\begin{center}
\Large{\bf OPTIMAL ASYMMETRIC CLONING MACHINE ON A GREAT CIRCLE
USING NO-SIGNALLING CONDITION}
\end{center}

\begin{center}
{{\bf Samir Kunkri}\( ^{1} \)}
, {{\bf Md. Manirul Ali}\( ^{2} \)}
, {{\bf Geetu Narang}\( ^{3} \)}
\ and {{\bf Debasis Sarkar}\( ^{4} \)}
\end{center}

\noindent{
\( ^{1 } \){\it Physics and Applied Mathematics Unit, Indian
Statistical Institute, 203 B.T. Road, Kolkata 700 108, India.}
E-mail: skunkri$_{-}$r@isical.ac.in}

\noindent{
\( ^{2 } \){\it S. N. Bose National Centre for Basic Sciences, JD
Block, Salt Lake City, Kolkata 700 098 , India.} E-mail:
mani@bose.res.in}

\noindent{
\( ^{3 } \){\it Department of Physics, Guru Nanak Dev University,
Amritsar 143005, India.} E-mail: g$_{-}$n29@yahoo.com}

\noindent{
\( ^{4 } \){\it Department of Applied Mathematics, University of
Calcutta, 92 A.P.C. Road, Kolkata 700 009, India.} E-mail:
debasis@cubmb.ernet.in}

\begin{abstract}
{\small {Assuming the condition of no superluminal signalling, we
got an upper bound on the quality of all asymmetric $ 1\rightarrow
2$ cloning machines, acting on qubits whose Bloch vectors lie on a
great circle. Then we constructed an $ 1\rightarrow 2$ cloning
machine, which asymmetrically clone all qubits corresponding to
this great circle, and this machine matches with that upper bound,
and hence this is optimal one.}}

\vspace{0.4cm} \noindent{PACS numbers: 03.67.Hk, 03.65.Bz,
89.70.+c}

\end{abstract}

\vspace{0.6cm}
 One can not exactly copy an arbitrarily given
quantum mechanical state \cite{nocloning}. However inexact quantum
cloning machine (QCM) exists. Bu\v{z}ek and Hillery have provided
a universal $1 \rightarrow 2$ cloning machine, which produces two
identical but imperfect copies of an arbitrary single qubit
\cite{bh}. Bru{\ss} et. al \cite{brussetal} have shown that this
symmetric (as the two copies produced are identical) universal
cloning machine of Bu\v{z}ek and  Hillery is optimal. Cerf
\cite{cerf} has provided a concept of universal asymmetric quantum
cloning when the two output states of the cloner are not identical
-- one is same as the original qubit after shrinkage of the Bloch
vector  by the factor $\eta_1$ and the other one is same as the
original qubit after shrinkage of the Bloch vector by the factor
$\eta_2$. A universal $ 1 \rightarrow 2 $ cloning network, for
asymmetric cloning, has been provided by Bu\v{z}ek et. al
\cite{bhb} using local unitary operations and controlled NOT
operations. And the symmetric optimal universal cloning machine of
Bu\v{z}ek and Hillery \cite{bh} has been reproduced. Allowing no-
signalling, Gisin \cite{gg} has reproduced the $1 \rightarrow 2 $
optimal universal symmetric cloning machine of Bu\v{z}ek and
Hillery \cite{bh} for arbitrarily given qubit. Ghosh et. al
\cite{gkr} have shown that the universal asymmetric cloning
machine provided by Bu\v{z}ek et. al \cite{bhb} is optimal one, by
using the constraint of no superluminal signalling.

 In this letter we derive an upper bound of the shrinking
factors for an optimal  asymmetric $1 \rightarrow 2$ (isotropic)
quantum cloning machine (OAQCMG) acting on all the qubits, whose
Bloch vector lie on a great circle, using the no-signalling
constraint, using the same procedure as used in references
\cite{gg} and \cite{gkr} (see \cite{comment}). Then we construct
an asymmetric $1 \rightarrow 2$ cloning machine for all the
qubits, mentioned above, whose fidelity matches with the above-
mentioned upper bound, and hence this is the optimal one. This
will automatically reproduce the result obtained by Bru{\ss} et.
al \cite{brussetal1} for optimal $1 \rightarrow 2$ symmetric QCM
acting on equatorial qubits.

 Let us take our great circle in the
x-z plane. Let
$$\rho^{in}_o(\overrightarrow{m}) =
\frac{1}{2}[I +  \overrightarrow{m}. \overrightarrow{\sigma}]=
\frac{1}{2}[I + m_x
\sigma_x + m_z \sigma_z]$$\\
be the original pure density matrix of the input single qubit,
entering into the OAQCMG, where $|\overrightarrow{m}| = 1$. We
want to clone (asymmetrically) this qubit universally for all
states on the great circle of the x-z plane ({\it i.e.},
independent of any Bloch vector $\overrightarrow{m}$ on the great
circle of the x-z plane), in such a way that the density matrices
of the two clones
at the output of the OAQCMG are of the forms\\
$$\rho^{out}_o(\overrightarrow{m}) =
Tr_b[\rho^{out}_{ob}(\overrightarrow{m})]= \frac{1}{2}[I +
\eta_1\overrightarrow{m}.\overrightarrow{\sigma}],$$\\
$$\rho^{out}_b(\overrightarrow{m})=
Tr_0[\rho^{out}_{ob}(\overrightarrow{m})] = \frac{1}{2}[I +
\eta_2\overrightarrow{m}. \overrightarrow{\sigma}],$$\\
where $o$ and $b$ correspond to the original system and the system
of the blank copy state, respectively and
$\rho^{out}_{ob}(\overrightarrow{m})$ be the two-qubit output
density matrix of the OAQCMG obtained after employing the trace
operation on the machine Hilbert space in the output pure state
$|\Psi (\overrightarrow{m})\rangle^{out}_{obM}$ (say) of original
qubit, blank copy and machine, obtained by applying the asymmetric
cloning operation on $\rho^{in}_o(\overrightarrow{m})$. Here the
reduction factors $\eta_1, \eta_2$ satisfy the condition $0 \le
\eta_1, \eta_2 \le 1$.
In full generality, $\rho^{out}_{ob}(\overrightarrow{m})$ can be
written as
\begin{equation}
\rho^{out}_{ob}(\overrightarrow{m}) = \frac{1}{4}\left[I\otimes I
+ \eta_1 (m_x \sigma_x\otimes I + m_z \sigma_z\otimes I)+ \eta_2
(m_xI\otimes  \sigma_x + m_z I \otimes \sigma_z)+\sum_{j,k=x,y,z}
t_{jk}\sigma_j\otimes\sigma_k\right],
\end{equation}
where $t_{jk}$'s are the real numbers. The OAQCMG will be
universal if it acts equally on all input states on the
above-mentioned great circle, {\it i.e.}, if
\begin{equation}
\rho^{out}_{ob}(R\overrightarrow{m})=
\rho^{out}_{ob}(\overrightarrow{m^{\prime}})= U(R) \otimes U(R)
\rho^{out}_{ob}(\overrightarrow{m}) U(R)^{\dagger} \otimes
U(R)^{\dagger},
\end{equation}
where $R = R(\overrightarrow{y},\beta)$ represents an arbitrary
rotation (in $SO(3)$) about y-axis through an angle $\beta$ of the
Bloch vector $\overrightarrow{m}$, and $U(R)=
e^{-\imath\frac{\beta}{2}\sigma_y}$ is the corresponding
$2\times2$ unitary operation (in $SU(2)$) acting on the two
dimensional Hilbert spaces corresponding to the two system $o$ and
$b$. Now from equation $(1)$ we may write the output density
matrix for $\overrightarrow{m^{\prime}} =
({m^{\prime}_x},0,{m^{\prime}_z})=R\overrightarrow{m}$ as,
\begin{equation}
\rho^{out}_{ob}(\overrightarrow{m}^{\prime}) =
\frac{1}{4}\left[I\otimes I + \eta_1 (m_x^{\prime} \sigma_x\otimes
I + m_z^{\prime}\sigma_z\otimes I)+ \eta_2 (m_x^{\prime}I\otimes
\sigma_x + m_z^{\prime} I \otimes \sigma_z)+\sum_{j,k=x,y,z}
t^{\prime}_{jk}\sigma_j\otimes\sigma_k\right]
\end{equation}\\
Then from equation (2) we have the following relations between the
correlation parameters $t_{jk}$ and $t^{\prime}_{jk}$ for
$j,k=x,y,z,$
\begin{equation}
\begin{array}{lcl}
t_{xx}^{\prime} &=& \cos^{2}\beta \ t_{xx} + \sin^{2}\beta \
t_{zz} + \sin\beta \cos\beta \ (t_{xz}+t_{zx}) \\
t_{xy}^{\prime} &=& \cos\beta \ t_{xy} + \sin\beta \ t_{zy}\\
t_{xz}^{\prime} &=& \sin\beta \cos\beta \ (t_{zz}-t_{xx}) +
\cos^{2}\beta \ t_{xz}- \sin^{2}\beta \ t_{zx}\\
t_{yx}^{\prime} &=& \cos\beta \ t_{yx} + \sin\beta \ t_{yz}\\
t_{yy}^{\prime} &=&  \ t_{yy} \\
t_{yz}^{\prime} &=& -\sin\beta \ t_{yx} + \cos\beta \ t_{yz}\\
t_{zx}^{\prime} &=& \sin\beta \cos\beta \ (t_{zz}-t_{xx}) -
\sin^{2}\beta \ t_{xz}+ \cos^{2}\beta \ t_{zx}\\
t_{zy}^{\prime} &=& -\sin\beta \ t_{xy} + \cos\beta \ t_{zy}\\
t_{zz}^{\prime} &=& -\sin\beta \cos\beta \ (t_{xz}+t_{zx}) +
\cos^{2}\beta \ t_{zz}+ \sin^{2}\beta \ t_{xx}\\
\end{array}
\end{equation}
Particularly, for $\beta =
0,\overrightarrow{m^{\prime}}=(0,0,1)=\uparrow$ (say), we have
from (4)$$ t_{xx}^{\prime} = \ t_{xx} ,\ t_{xy}^{\prime} =  \
t_{xy} ,\ t_{xz}^{\prime} =  \ t_{xz} ,\ t_{yx}^{\prime} =  \
t_{yx} ,\
 t_{yy}^{\prime} =  \ t_{yy} $$
 $$ t_{yz}^{\prime} =  \ t_{yz} ,\
 t_{zx}^{\prime} =  \ t_{zx} ,\ t_{zy}^{\prime} =  \ t_{zy}, \
  t_{zz}^{\prime} =  \ t_{zz}$$
for $ \beta =
180^o,\overrightarrow{m^{\prime}}=(0,0,-1)=\downarrow$
 (say), $$t_{xx}^{\prime} =  \ t_{xx} ,\ t_{xy}^{\prime} =  \ -t_{xy} ,\
t_{xz}^{\prime} =  \ t_{xz} ,\ t_{yx}^{\prime} =  \ -t_{yx} ,\
 t_{yy}^{\prime} =  \ t_{yy} $$
 $$ t_{yz}^{\prime} =  \ -t_{yz} ,\
 t_{zx}^{\prime} =  \ t_{zx} ,\ t_{zy}^{\prime} =  \ t_{zy}, \
  t_{zz}^{\prime} =  \ t_{zz}$$
for $ \beta =
90^o,\overrightarrow{m^{\prime}}=(1,0,0)=\longrightarrow$
 (say), $$t_{xx}^{\prime} =  \ t_{zz} ,\ t_{xy}^{\prime} =  \ t_{zy} ,\
t_{xz}^{\prime} =  \ -t_{zx} ,\ t_{yx}^{\prime} =  \ t_{yz} ,\
 t_{yy}^{\prime} =  \ t_{yy} $$
 $$ t_{yz}^{\prime} =  \ -t_{yx} ,\
 t_{zx}^{\prime} =  \ -t_{xz} ,\ t_{zy}^{\prime} =  \ -t_{xy}, \
  t_{zz}^{\prime} =  \ t_{xx}$$
for $ \beta =
270^o,\overrightarrow{m^{\prime}}=(-1,0,0)=\longleftarrow$
 (say), $$t_{xx}^{\prime} =  \ t_{zz} ,\ t_{xy}^{\prime} =  \ -t_{zy} ,\
t_{xz}^{\prime} =  \ -t_{zx} ,\ t_{yx}^{\prime} =  \ -t_{yz} ,\
 t_{yy}^{\prime} =  \ t_{yy} $$
 $$ t_{yz}^{\prime} =  \ t_{yx} ,\
 t_{zx}^{\prime} =  \ -t_{xz} ,\ t_{zy}^{\prime} =  \ t_{xy}, \
  t_{zz}^{\prime} =  \ t_{xx}$$Now, the no signaling condition
  implies \cite{gisin}:
  \begin{equation}
  \rho^{out}_{ob}(\uparrow)+\rho^{out}_{ob}(\downarrow)=\rho^{out}_{ob}(\rightarrow)+
\rho^{out}_{ob}(\leftarrow).
\end{equation}
Which provides us,
$$t_{xx}=t_{zz}, \ t_{xz}=-t_{zx}.$$
Thus,
$$
\rho^{out}_{ob}(\uparrow) = \frac{1}{4}[ I \otimes I + \eta_1
\sigma_z\otimes I +  \eta_2 I\otimes \sigma_z +
t_{xx}(\sigma_x\otimes\sigma_x+\sigma_z\otimes\sigma_z)$$
$$+ t_{yy} \ \sigma_y\otimes\sigma_y+t_{xy}\
\sigma_x\otimes\sigma_y
+t_{xz}(\sigma_x\otimes\sigma_z-\sigma_z\otimes\sigma_x)
+t_{yx}\sigma_y\otimes\sigma_x+t_{yz}\sigma_y\otimes\sigma_z
+t_{zy}\sigma_z\otimes\sigma_y ]$$

$$= \frac{1}{4} \times$$
$$\left[\begin{array}{cccc}
(1+\eta_1+\eta_2+t_{xx})&-(t_{xz}+\it{i}t_{zy})&(t_{xz}-\it{i}t_{yz})&(t_{xx}-t_{yy})-\it{i}(t_{xy}+t_{yx})\\
(-t_{xz}+\it{i}t_{zy})&(1+\eta_1-\eta_2-t_{xx})&(t_{xx}+t_{yy})+\it{i}(t_{xy}-t_{yx})&(-t_{xz}+\it{i}t_{yz})\\
(t_{xz}+\it{i}t_{yz})&(t_{xx}+t_{yy})-\it{i}(t_{xy}-t_{yx})&(1-\eta_1+\eta_2-t_{xx})&(t_{xz}+\it{i}t_{zy})\\
(t_{xx}-t_{yy})+\it{i}(t_{xy}+t_{yx})&-(t_{xz}+\it{i}t_{yz})&(t_{xz}-\it{i}t_{zy})&(1-\eta_1-\eta_2+t_{xx})\end{array}\right].
$$\\ Since $\rho^{out}_{ob}(\uparrow)$ must be positive semi-definite,
we must have,\\
\begin{equation}
\eta_{1}^{2} + \eta_{2}^{2} \leq 1 -
t_{yy}^{2}-t_{xy}^{2}-t_{yx}^{2}-t_{yz}^{2}-t_{zy}^{2}.
\end{equation}
From equation (6) we see that maximum allowed values of both
$\eta_{1}$  and  $\eta_{2}$ will occur when
$t_{yy}=t_{xy}=t_{yx}=t_{yz}=t_{zy}=0$, hence
\begin{equation}
\label{optimalcurve} {\eta_{1}}^{2} + {\eta_{2}}^{2}= 1.
\end{equation}
For the symmetric case, ${\eta_{1}} = {\eta_{2}} = {\eta_{max}}$
(say), then
$${\eta_{max}}=\frac{1}{\sqrt{2}}.$$
The optimal fidelity for symmetric case is then
$$
F^{opt} = \frac{1}{2}(1+\eta_{max})= \frac{1}{2}+
\sqrt{\frac{1}{8}},$$ which is exactly the same result obtained by
Bru{\ss} et. al \cite{brussetal1} for optimal $1 \rightarrow 2$
symmetric cloning for all qubits lie on the great circle of the
x-z plane.
Interestingly, for all the qubits lying on the great circle of the
x-z plane, the allowed maximum values of $\eta_{1}$ and $\eta_{2}$
lie on a circle (see fig. 1), where as in the case of universal
optimal asymmetric cloning machine, it is on an ellipse
\cite{gkr}.

\begin{center}
\unitlength=0.5mm
\begin{picture}(180,180)(0,0)
\put(0,55){\vector(1,0){110}}
\put(115,55){\makebox(0,0){{\tiny{$\eta_1$}}}}
\put(60,0){\vector(0,1){110}}
\put(60,115){\makebox(0,0){{\tiny{$\eta_2$}}}}
\put(60,55){\circle*{2}}
\put(60,55){\circle{80}} \put(74,55){\circle*{2}}
\put(60,69){\circle*{2}} \put(41,36){\makebox(0,0){\tiny{(0, 0)}}}
\put(45,40){\vector(1,1){13}} \put(88,36){\makebox(0,0){\tiny{(1,
0)}}} \put(88,40){\vector(-1,1){12}}
\put(35,69){\makebox(0,0){\tiny{(0, 1)}}}
\put(42,70){\vector(1,0){13}}
\put(73,68){\makebox(0,0){{\tiny{${\cal C}$}}}}
\end{picture}
\end{center}


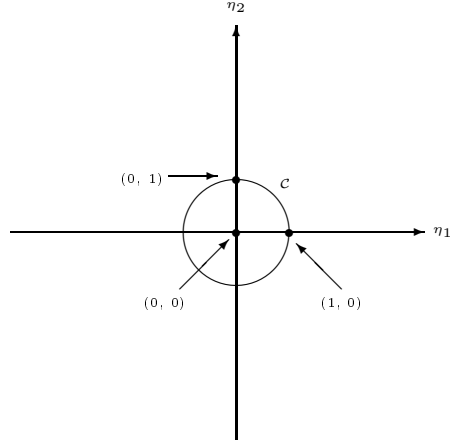
{\noindent{\small{{\bf Figure 1} : The part ${\cal C} = \{(\eta_1,
\eta_2) \in {I\!\!R}^2  : 0 \le \eta_1, \eta_2 \le 1,~ \eta_1^2 +
\eta_2^2 = 1\}$ of the unit circle $\eta_1^2 + \eta_2^2 = 1$
represents the maximum allowed values of the reduction factors
$\eta_1, \eta_2$ of asymmetric (isotropic) $1 \rightarrow 2$
cloning machines of all the qubits whose Bloch vectors lie on any
given great circle.}}}

\vspace{0.5cm}

 Thus we get an upper bound (given by equation (7)) on the quality of
 asymmetrically $1 \rightarrow 2$ cloning of qubits, whose Bloch
 vectors lie on the great circle in x-z plane, provided we take it
 to be granted that no faster than light signalling is possible here
 - an assumption, of which we have, still today, no exception \cite{shimony}.

 Next we consider the following unitary transformation which would
 give rise to an asymmetric $1 \rightarrow 2$ cloing machine for
 all qubits, whose Bloch vectors lie on this great circle in the
 x-z plane :
 \begin{equation}
 \label{oac}
 \begin{array}{lcl}
 U\left(| 0 \rangle_{o} \otimes | . \rangle_{b}\otimes | m \rangle_{M}\right) &=& \left(A | 00 \rangle_{ob} +
 D | 11 \rangle_{ob} \right) \otimes | 0 \rangle_{M} + \left(B | 01 \rangle_{ob} +  C | 10 \rangle_{ob}\right) \otimes | 1 \rangle_{M},\\
U\left(| 1 \rangle_{o} \otimes | . \rangle_{b}\otimes | m
\rangle_{M}\right) &=& \left(A | 11 \rangle_{ob} +
 D | 00 \rangle_{ob} \right) \otimes | 1 \rangle_{M} + \left(B | 10 \rangle_{ob} +  C | 01 \rangle_{ob}\right) \otimes | 0 \rangle_{M},\\
\end{array}
\end{equation}
where
\begin{equation}
\label{coeff}
\begin{array}{lcl}
A &=& \frac{1}{2}\left\{\left(1+\eta_{1}\right) \left(1+\eta_{2}\right)\right\}^{1/2},\\
B &=& \frac{1}{2}\left\{\left(1+\eta_{1}\right) \left(1-\eta_{2}\right)\right\}^{1/2},\\
C &=& \frac{1}{2}\left\{\left(1-\eta_{1}\right) \left(1+\eta_{2}\right)\right\}^{1/2},\\
D &=& \frac{1}{2}\left\{\left(1-\eta_{1}\right)
\left(1-\eta_{2}\right)\right\}^{1/2},
\end{array}
\end{equation}
and where $|.\rangle_{b}$ being a fixed blank state, $| m
\rangle_{M}$ being a fixed machine state. Here $0 \le \eta_1,
\eta_2 \le 1$. It is obvious that the transformation $U$, given in
(\ref{oac}), is unitary. Let $|\psi\rangle_o = \alpha |0\rangle_o
+ \beta |1\rangle_o$ be any state whose Bloch vector lies on the
great circle of x-z plane, and so $\alpha$, $\beta$ are real
numbers with $\alpha^2 + \beta^2 = 1$. It can be shown that
\begin{equation}
\label{isotropycond}
\begin{array}{lcl}
Tr_{bM} \left(P\left[U\left(|\psi\rangle_o \otimes |.\rangle_b
\otimes |m\rangle_M\right)\right]\right) &=& s_1
|\psi\rangle_o\langle\psi| + \frac{1 - s_1}{2} I_o,\\
Tr_{oM} \left(P\left[U\left(|\psi\rangle_o \otimes |.\rangle_b
\otimes |m\rangle_M\right)\right]\right) &=& s_2
|\psi\rangle_b\langle\psi| + \frac{1 - s_2}{2} I_b,
\end{array}
\end{equation}
for some $s_1$, $s_2$ (with $0 \le s_1, s_2 \le 1$; $I_o, I_b$
being identity operators of the systems $o$, $b$ respectively),
which are independent of $|\psi\rangle$, provided $s_1 = \eta_1$,
$s_2 = \eta_2$, and $\eta_1^2 + \eta_2^2 = 1$ ({\it i.e.},
condition (\ref{optimalcurve}) is satisfied). Thus we see that
(when the constraint of no superluminal signalling is imposed on
the principles of quantum mechanics) the unitary transformation
$U$ in (\ref{oac}) gives rise to the optimal asymmetric $1
\rightarrow 2$ cloning machine of all the qubits, whose Bloch
vectors lie on the great circle of the x-z plane, provided the
coefficients $A$, $B$, $C$, $D$ satisfy the relations given in
(\ref{coeff}), and where the reduction factors $\eta_1$, $\eta_2$
lie on the curve (\ref{optimalcurve}) \cite{clarification}.

Therefore, the shrinking factors $ \eta_1 , \eta_2 $,
corresponding to OACQMG,
satisfy the relation $ {\eta_1}^2 + {\eta_2}^2 = 1$, provided the
constraint of no superluminal signalling is taken into account.
The optimal symmetric $1 \rightarrow 2$ cloning machine of all the
qubits whose Bloch vectors lie on any great circle (given by
Bru{\ss} et. al \cite{brussetal1}) is a special case of (7), where
$ \eta_1 = \eta_2 = \frac{1}{\sqrt{2}}$.

In summary, we got the optimal asymmetric $ 1 \rightarrow 2 $
cloning machine of all the qubits whose Bloch vectors lie on a
great circle, using the assumption of no faster than light
signalling. Although it is not clear whether the upper bound on
the quality (in terms of fidelity) of universal (or,
set-dependent) isotropic $1 \rightarrow 2$ quantum cloning machine
(derived from no-signalling conditions) matches with the fidelity
of the optimal universal (or, set-dependent) isotropic  $1
\rightarrow 2$ quantum cloning machine (derived from quantum
mechanical principles), for systems in higher dimension
\cite{brussetal2}, in two dimension, these two processes match
exactly ({\it i.e.}, the fidelities of cloning machines, one
derived from no-signalling constraints and another one derived
from the principles of quantum mechanics, match exactly), for
universal set as well as for the set of all qubits whose Bloch
vectors lie on a great circle. It may be noted that unlike the
case of optimal universal $1 \rightarrow 2$ cloning of qubits, the
joint output state $\rho_{ob} (\psi) \equiv Tr_{M}
\left(P\left[U\left(|\psi\rangle_o \otimes |.\rangle_b \otimes
|m\rangle_M\right)\right]\right)$, after applying the optimal
cloning machine, given by (\ref{oac}), is separable, for every
input state $|\psi\rangle$ whose Bloch vector lies on the great
circle of the x-z plane.

\vspace{0.5cm} {\noindent {{\large{\bf Acknowledgement}}: The
authors thank Sibasish Ghosh, Guruprasad Kar and Anirban Roy for
helpful discussions. S. K. and M. M. A. each, acknowledges partial
support by the Council of Scientific and Industrial Research,
Government of India, New Delhi. M. M. A. thanks Archan S. Majumdar
for  encouragement to carry out this work. G. N. acknowledges the
support of Indian Statistical Institute, while visiting the
Physics and Applied Mathematics Unit of this Institute, during
which a major part of this work has been done.}}


\begin{thebibliography}{99}
\bibitem{nocloning}G. C. Ghirardi, in a referee report of an
artical submitted in {\it Found. Phys.} (1981); W. K. Wootters and
W. H. Zurek, {\it Nature} {\bf 299}, 802 (1982); D. Dieks, {\it
Phys. Lett. A} {\bf 92}, 271 (1982); H. Yuen, {\it Phys. Lett. A}
{\bf 113}, 405 (1986).
\bibitem{bh}V. Bu\v{z}ek and M. Hillery, {\it Phys. Rev. A} {\bf
54}, 1844 (1996).
\bibitem{brussetal}D. Bru{\ss}, D. P. DiVincenzo, A. Ekert, C. A. Fuchs, C. Macchiavello, and J. A. Smolin, {\it Phys. Rev. A} {\bf
57}, 2368 (1998).
\bibitem{cerf}N. J Cerf, {\it ``Quantum cloning and the capacity of the Pauli Channel"}, quant-ph/9803058; N. J. Cerf, {\it Acta Physica Slovaca} {\bf
48}, 115 (1998).
\bibitem{bhb}V. Bu\v{z}ek, M. Hillery, and R. Bednik, {\it Acta
Physica Slovaca} {\bf 48}, 177 (1998).
\bibitem{gg}N. Gisin, {\it Phys. Lett. A} {\bf 242}, 1 (1998).
\bibitem{gkr}S. Ghosh, G. Kar, and A. Roy, {\it Phys. Lett. A} {\bf 261},
17 (1999).
\bibitem{comment}From now on we will use only isotropic cloning
machines.
\bibitem{brussetal1}D. Bru{\ss}, M. Cinchetti, G. M. D'Ariano, and C. Macchiavello, {\it Phys. Rev. A} {\bf 62}, 12302 (2000).
\bibitem{gisin}The no signalling condition imposes that the
mixtures of output states corresponding to indistinguishable
mixtures of input states are themselves indistinguishable. This is
more general than the condition used here. However, it turns out
that in the present context, the simple condition $(5)$ is
sufficient (see ref. \cite{gg} and \cite{gkr}).
\bibitem{shimony}Actually, according to Shimony [A. Shimony, in
{\it Foundations of Quantum Mechanics in the Light of New
Technology}, ed. S. Kamefuchi, {\it Phys. Soc. Japan}, Tokyo,
1983], there is a peaceful coexistence between quantum mechanics
and relativity.
\bibitem{clarification}Note that in equations (\ref{oac}) and (\ref{coeff}),
if $\eta_1^2 + \eta_2^2 < 1$, the unitary operation $U$ (in
(\ref{oac})) does not give rise to a non-optimal asymmetric $1
\rightarrow 2$ cloning machine of qubits on the great circle of
the x-z plane, for some suitable values of the shrinking factors
$s_1, s_2$ (each of which would depend on $\eta_1, \eta_2$).
\bibitem{brussetal2}D. Bru{\ss}, G. M. D'Ariano, C. Macchiavello, and M. F. Sacchi, {\it ``Approximate quantum cloning and the impossibility of superluminal information transfer" }, quant-ph/0010070.
\end{thebibliography}
\end{document}